\documentclass[copyright]{eptcs}
\providecommand{\event}{GaM 2015}

\usepackage{amsmath,latexsym,amsfonts,amssymb,amsthm,verbatim}
\usepackage{rotating,graphs,stmaryrd}
\usepackage{subfig,pgf,xcolor,tikz}
\usepackage{cite}
\usepackage{breakurl}

\theoremstyle{remark}
\newtheorem{example}{Example}

\newcommand{\ttt}{\texttt}

\newcommand{\mtt}{\mathtt}

\newcommand{\tuple}[1]{\langle#1\rangle}

\usetikzlibrary{arrows,fit,backgrounds}
\tikzset{arrowin/.style={<-,>=latex,semithick},arrowout/.style={->,>=latex,semithick},root/.style={circle,draw,line width=2pt},box/.style={rectangle, rounded corners, minimum width=1cm, minimum height=1cm,text centered, draw=black}}

\begin{document}

\title{A Reference Interpreter for the Graph\\ Programming Language GP 2}

\author{Christopher Bak\thanks{The first two authors are supported by grants from the Engineering and Physical Sciences Research Council in the UK.},\hspace{.2em} Glyn Faulkner\footnotemark[1],\hspace{.2em} Detlef Plump and Colin Runciman
\institute{Department of Computer Science, The University of York, UK}
}

\def\authorrunning{C. Bak, G. Faulkner, D. Plump \& C. Runciman} 
\def\titlerunning{A Reference Interpreter for GP~2}

\maketitle
\thispagestyle{empty}

\begin{abstract}
GP~2 is an experimental programming language
for computing by graph transformation.
An initial interpreter for GP~2, written
in the functional language Haskell, provides
a concise and simply structured reference
implementation.
Despite its simplicity, the performance of
the interpreter is sufficient for the
comparative investigation of a range of test
programs.
It also provides a platform for the development
of more sophisticated implementations.
\end{abstract}

\section{Introduction}

GP~2 is an experimental programming language in which the major part of
the computational state is a labelled directed graph, and the basic
units by which computational progress is made are subgraph-replacement
rules.
Choices of rules and subgraphs are non-deterministic, and some of
the control structures above the level of rules involve back-tracking.

The implementation of such a programming language poses some
interesting challenges and opportunities.
Our ultimate goal is to produce a compiler from GP~2 to
high-performance executable code.
This paper reports a first stage towards that goal, the development
of a \emph{reference interpreter} for GP~2.
By this we mean an interpreter written with the main aim of
being clear, concise and correct.
Where there are design choices, simplicity of
definition takes priority over other considerations
such as performance and the richness of functionality. The interpreter contains only around 1,000 lines of Haskell source code. Even so, we shall show that it is usable in practice.

Section~\ref{sec:graph-programs} outlines and illustrates the 
graph programming language GP~2.
Section~\ref{sec:benchmark} presents a small set of test programs
written in GP~2.
Section~\ref{sec:usesrequirements} considers the expected uses of
a reference interpreter, and consequent requirements.
Section~\ref{sec:implementation} describes our reference interpreter for
GP~2.
Section~\ref{sec:performanceevaluation} sets out the measured results of using the reference
interpreter to evaluate test programs.
Section~\ref{sec:relatedandfuture} briefly discusses related work and
indicates some of our own expected lines of future work.
Section~\ref{sec:conclusions} draws overall conclusions from our work
on the reference interpreter for GP~2.

\section{Graph Programs}
\label{sec:graph-programs}

This paper focusses on GP~2, a successor to the graph programming language GP \cite{Plump09a,Plump12a}.
GP is a domain-specific language which aims to support formal reasoning on graph programs (see \cite{Poskitt-Plump12a} for a Hoare-logic approach to verifying GP programs). We give a brief introduction to GP~2, mainly by example. The definition of the language, including a formal operational semantics, can be found in \cite{Plump12a}. 

A graph program consists of declarations of conditional graph transformation rules and macros, and exactly one main command sequence. Graphs are directed and may contain  loops and parallel edges. The rules operate on a \emph{host graph}\/ (or input graph) whose nodes and edges are labelled with a list of integers and character strings. Besides the list, a label may contain a \emph{mark}\/ which is one of the values \ttt{red}, \ttt{green}, \ttt{blue}, \ttt{grey} and \ttt{dashed} (where \ttt{grey} and \ttt{dashed} are reserved for nodes and edges, respectively). For example, the node label on the right-hand side of the rule \ttt{init} in Figure \ref{fig:vertex-colouring} is the pair $\tuple{\mtt{x{:}1},\, \mtt{grey}}$.

Variables in rules are of type \texttt{int}, \texttt{char}, \texttt{string}, \texttt{atom} or \texttt{list}, where \texttt{atom} is the union of \texttt{int} and \texttt{string}. Atoms are considered as lists of length one, hence integers and strings are also lists. Similarly, characters are considered as strings of length one. Given lists $\mtt{x}$ and $\mtt{y}$, their concatenation is written \texttt{x:y} (not to be confused with the list-cons operator in Haskell). 

\begin{example}[Transitive Closure]
The principal programming constructs in GP~2 are conditional graph-transformation rules labelled with expressions. The program in Figure \ref{fig:transitive-closure} applies the single rule \ttt{link} \emph{as long as possible} to a host graph. In general, any subprogram can be iterated with the postfix operator ``\ttt{!}''. (A composite loop $(P_1\mtt{;}\dots\mtt{;}P_n)\mtt{!}$ terminates if any of the components $P_i$\/ \emph{fails}, meaning that some rule in $P_i$\/ could not be matched. In this case the loop finishes with the graph on which the current iteration of the body $(P_1;\dots;P_n)$ was entered. See \cite{Plump12a} for details.)

\begin{figure}[htb]
\begin{center}
 \input{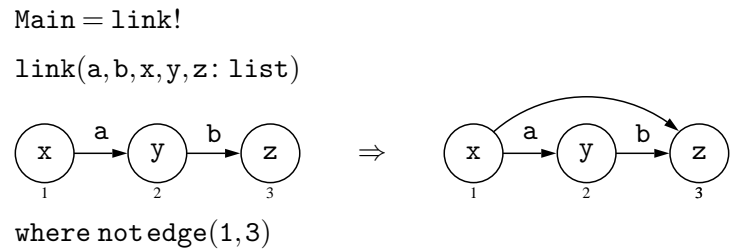}
\end{center}
\caption{Program for transitive closure}\label{fig:transitive-closure}
\end{figure}

Applying \ttt{link} amounts to non-deterministically selecting a subgraph of the host graph that matches \ttt{link}'s left graph, and adding to it an edge from node 1 to node 3 provided there is no such edge (with any label). The application condition ensures that the program terminates and extends the host graph with a minimal number of edges. Rule matching is injective and involves instantiating variables with concrete values (see also below).

A graph is \emph{transitive} if for each directed path from a node $v$ to another node $v'$, there is an edge from $v$ to $v'$.  Given any graph $G$, the program in Figure \ref{fig:transitive-closure} produces the smallest transitive graph that results from adding unlabelled edges to $G$.\footnote{``Unlabelled'' edges are actually labelled with the empty list.} This graph is unique up to isomorphism and requires at most $n^2$ applications of \ttt{link}, where $n$\/ is the number of nodes in $G$. \qed
\end{example}

\begin{figure}[htb]
\begin{center}
 \input{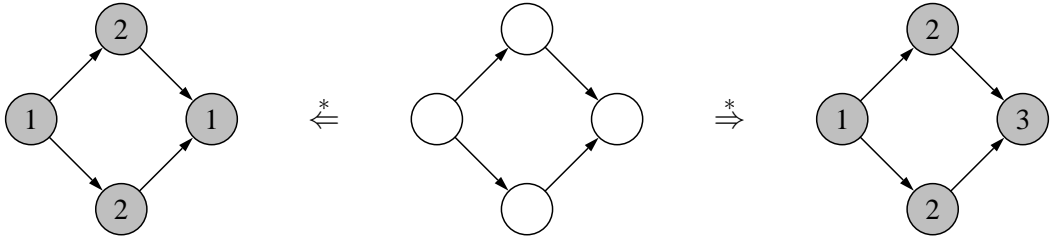}
\end{center}
\caption{Different results from vertex colouring}\label{fig:colour_results}
\end{figure}

\begin{example}[Vertex Colouring]
The program in Figure \ref{fig:vertex-colouring} assigns a \emph{colour}\/ to each node of the host graph, such that non-loop edges have differently coloured endpoints. Positive integers are used as colours because, in general, an unbounded number of colours is needed. The program replaces each node label $l$\/ with $l{:}i$, where $i$\/ is the node's colour. In addition, the rule \ttt{init} shades nodes to prevent repeated application to the same node.

\begin{figure}[htb]
\begin{center}
 \input{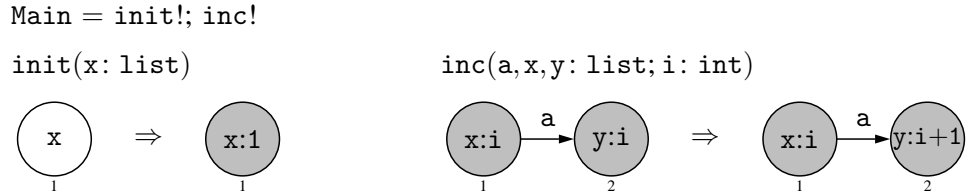}
\end{center}
\caption{Program for vertex colouring}\label{fig:vertex-colouring}
\end{figure}

Rule \ttt{inc} is applied to the host graph as long as there are edges with identically coloured endpoints. It can can be shown that this terminates after at most $n^2$ rule applications, where $n$\/ is the number of nodes. In contrast to the previous example program, \emph{different graphs may result}\/ from this process. In particular, there is no guarantee that the number of colours produced is minimal. For instance, Figure \ref{fig:colour_results} shows two different colourings produced for the same host graph.
\qed
\end{example}

\vspace{.5\baselineskip}
\noindent
\emph{Other program constructs.}
A GP~2 command not used in the example programs is a rule set $\mtt{\{}r_1,\dots,r_n\mtt{\}}$. This command non-deterministically applies any of the rules to the current host graph. The application \emph{fails}\/ if none of the left-hand graphs in the rules matches a subgraph. Matches must be injective and are only valid if they do not result in \emph{dangling edges}. (More formally, GP~2 is based on the double-pushout approach with injective matching, extended with relabelling and rule schemata \cite{Plump12a}.)

Another construct not yet discussed is the branching command \ttt{if} $C$ \ttt{then} $P$ \ttt{else} $Q$, where $C$, $P$ and $Q$ are arbitrary command sequences. This is executed on a host graph $G$ by first executing $C$ on a copy of $G$. If $C$ succeeds, $P$\/ is executed on the original graph $G$; otherwise, $Q$ is executed on $G$. The command \ttt{try} $C$ \ttt{then} $P$ \ttt{else} $Q$ has a similar effect, except that $P$\/ is executed on the graph resulting from $C$'s execution. 

\section{Benchmark Programs}
\label{sec:benchmark}
 
We envisage GP 2 as a general-purpose language for graph problems, hence the reference interpreter should be tested on algorithms of varying complexity. This is different from the benchmarking reported in \cite{Varro-Schuerr-Varro05a} which focusses on a deterministic program with very limited complexity. In Section \ref{sec:performanceevaluation}, we evaluate the performance of our interpreter on a small set of benchmark programs. These include the programs for transitive closure and vertex colouring, and three more programs which we describe in this section.

\begin{figure}[t]
\begin{center}
\input{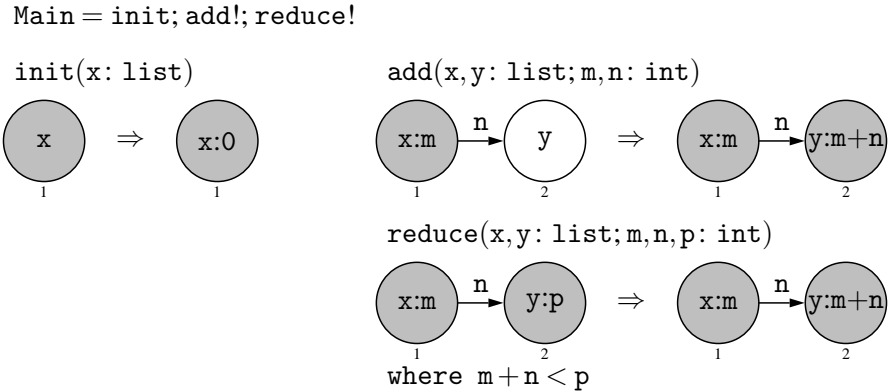}
\end{center}
\caption{Program for shortest distances}\label{fig:shortest-distances}
\end{figure}

\vspace{.5\baselineskip}
\noindent
\emph{Shortest distances.} The program in Figure \ref{fig:shortest-distances} expects an input graph $G$\/ containing a unique grey node $s$, where edge labels are assumed to be non-negative integers. A unique output graph is obtained by marking grey each node reachable from $s$ and replacing its label $l$\/ with $l{:}d$, where $d$\/ is the shortest distance from $s$. (A distance is the sum of the edge labels of a directed path.)
  
The program first assigns distance 0 to the unique start node $s$. Then the loop \ttt{add!} traverses the nodes reachable from $s$, assigning distances by adding edge labels. In a second phase, the loop \ttt{reduce!} minimizes distances by searching for edges whose sum of source node distance and edge label is smaller than the target node distance, and replacing the target node distance with the sum.

The requirement that edge labels are non-negative ensures that the program terminates. It can be relaxed by allowing negative edge labels but requiring that directed cycles have a non-negative overall distance.

\begin{figure}[t]
\begin{center}
\input{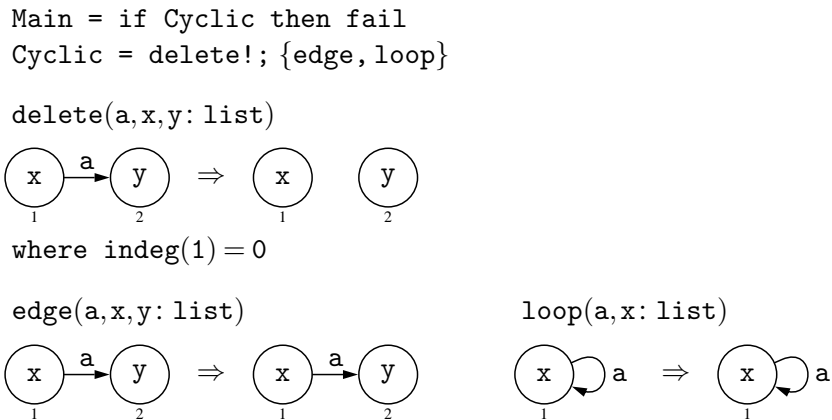}
\end{center}
\caption{Program for recognising acyclic graphs}\label{fig:acyclicity}
\end{figure}
\vspace{.5\baselineskip}
\noindent
\emph{Recognising acyclic graphs.} The program in Figure \ref{fig:acyclicity} checks whether its input graph is acyclic. If this is the case, the program preserves its input graph, otherwise it fails. Suppose we call the program \ttt{acyclic} to use it as a macro in the program \ttt{if} \ttt{acyclic} \ttt{then} $P$ \ttt{else} $Q$. Given any input graph $G$, this program will test whether $G$\/ is acyclic and, depending on the result, either execute $P$ or $Q$ on $G$.  
  
The presence of cycles is checked by deleting as long as possible edges whose sources have no incoming edges, and testing whether any edges remain. This is correct since an application of \ttt{delete} preserves both the absence and the presence of cycles (by the condition of the rule). Moreover, a graph to which \ttt{delete} is not applicable is acyclic if and only if it is edge-less (every acyclic graph with edges must contain an edge to which \texttt{delete} is applicable). 

%
%
%

\begin{figure}[p]
\begin{center}
\input{Programs/sierpinski.prog}
\end{center}
\caption{Program for generating Sierpinski triangles}\label{fig:Sierpinski-program}
 \begin{center}
  \includegraphics[scale=.35,angle=-15]{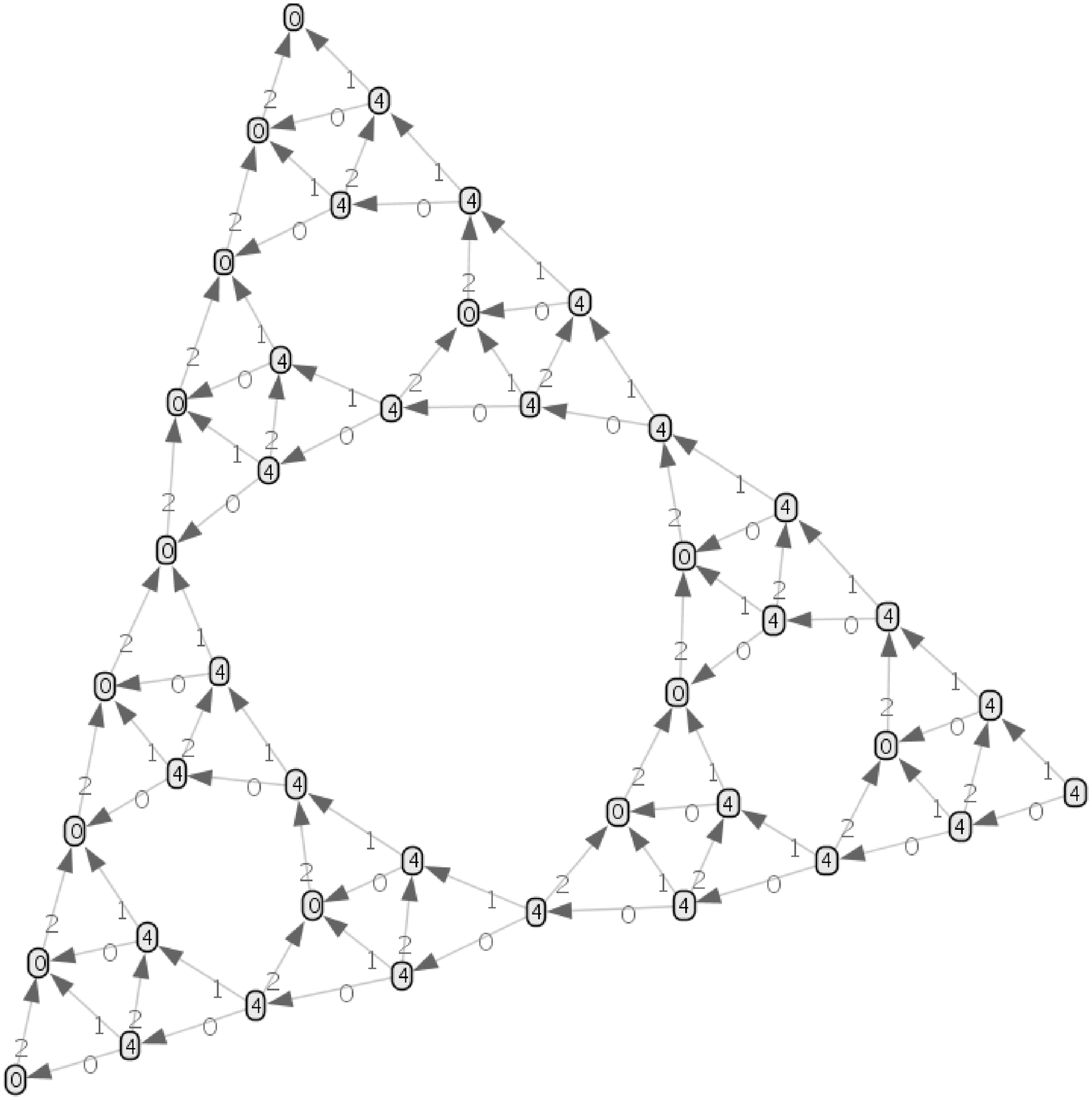}
 \end{center}
\vspace*{-2.5cm}
\caption{Third generation Sierpinski triangle \label{fig:sierpinski}}
\end{figure}

\vspace{.5\baselineskip}
\noindent
\emph{Generating Sierpinski triangles.} A \emph{Sierpinski triangle} is a self-similar geometric structure which can be recursively defined. Figure \ref{fig:sierpinski} shows a Sierpinski triangle of generation three, composed of three second-generation triangles, each of which consists of three triangles of generation one.\footnote{The geometric layout was created by the graphical interface of the GP 1 implementation \cite{Manning-Plump08b}.}

The program in Figure \ref{fig:Sierpinski-program} expects as input a single node labelled with the generation number of the Sierpinski triangle to be produced. The rule \texttt{init} creates the Sierpinski triangle of generation 0 and turns the input node into a ``control node'' with label $x{:}0$, holding the required generation number $x$ together with the current generation number.

After initialisation, the nested loop $\mtt{(inc;\, expand!)!}$ is executed. In each iteration of the outer loop, \texttt{inc} increases the current generation number if it is smaller than the required number (which is checked by the rule's condition). If the test is successful, the inner loop \ttt{expand!} performs a Sierpinski step on each triangle whose top node is labelled with the current generation number: the triangle is replaced by four triangles such that the top nodes of the three outer triangles are labelled with the next higher generation number. The test $\mathtt{x > y}$ fails when the required generation number has been reached. In this case the application of \texttt{inc} fails, causing the outer loop to terminate and return the current graph which is the Sierpinski triangle of the requested generation.

Sierpinski triangles pose a hard challenge for graph transformation: generating the $n$-th triangle requires space and a number of rule applications exponential in $n$. This problem was part of the 2007 tool contest for graph transformation, where the goal was to generate triangles of generation numbers as high as possible and as fast as possible \cite{Taentzer_et_al08a}.

\section{Reference Interpreters: Uses and Requirements}
\label{sec:usesrequirements}

A reference interpreter for a new programming language such as GP2
has several potential uses.
Each has consequences for the
way the reference interpreter is written and the
facilities it provides.

\vspace{.5\baselineskip}
\noindent
\emph{An arbiter for programmers.}
A programmer working in a new language needs to
know whether what they are writing is a valid
program, and whether the effect of executing it
is the effect they intend.
To resolve such issues, the programmer may want to
use a reference interpreter as a black box,
checking the output it produces given their
program as input.
Or they may wish to look at a salient part of the
source-code for the interpreter, to confirm some
aspect of the language they are unsure about.

It follows that a reference interpreter should
provide as output at least a report whether a
program is valid, and if so a clear representation
of the result when it is evaluated.
It also follows that the source-code for a
reference interpreter should
be organised in such a way that salient components
are easy to identify.
For ease of reading it should be written using a
consistent style in a modest subset of a suitable
high-level language.

\vspace{.5\baselineskip}
\noindent
\emph{An arbiter for implementors.}
An implementer of a programming language,
developing their own interpreter or compiler,
needs a standard against which to test the correctness
of their implementation.
There are two main respects in which any
implementation should agree with a reference interpreter
as a defining standard.
They should agree which programs are valid,
and for valid programs they should agree the results
of executing them.
Like application programmers, implementers too may
wish sometimes to use the reference interpreter as
a black box, but at other times to consult its
internal definitions. 

There are additional requirements for this use,
bearing in mind the likely development or generation
of many test programs.
The representation of the
reference interpreter's results for such programs
should be amenable to automated comparison.
This comparison presents particular challenges in GP~2 since
behaviour of programs may be non-deterministic,
or programs may not terminate, or both.
The number of test programs may be large
--- there may even be arbitrarily many test programs generated dynamically.
So although performance is not a design goal for the reference
interpreter, its performance should be good enough to
make such multi-test comparisons feasible.

\vspace{.5\baselineskip}
\noindent
\emph{A prototype for application developers.}
If no production compiler has been developed for the language,
or none is yet available to an application developer,
they may need to use a reference interpreter as
an initial development platform.

During the development of application programs, errors
are common.
So, for this use, a reference interpreter should provide
not only a check for valid programs, but a rapid check
with informative reports of errors.
Yet elaborate error handling must not obscure the
definitional style in which the interpreter is written.
Similarly, it is desirable to have the option of some
kind of trace or other informative report to shed
light on failures or unexpected results when a program
is evaluated.
Here again, the machinery must not obscure the basic
definitions for evaluation, nor should it impose heavy
performance costs when performance of the interpreter
has already been sacrificed in favour of simplicity.

\vspace{.5\baselineskip}
\noindent
\emph{A prototype for implementation developers.}
As well as using a reference interpreter to verify correctness,
implementers may wish to use it as the starting point in the
development of another interpreter or a compiler.
The whole course of such a development might even be defined as
the successive replacement of interpreter components by
alternatives giving higher performance, or richer information,
at the cost of greater complexity.
The advantage of this approach is that as each replacement
is introduced it can be checked as a new component in an already
tried system.

This use of a reference interpreter requires a
modular design with simple and clearly defined interfaces
between components.
Concerns should be separated so far as possible, avoiding
dependencies that are not strictly necessary.
Options for development by successive replacement may be further
increased by choosing a host programming system for the reference
interpreter that has a well-developed foreign-language interface.

\section{Implementation}
\label{sec:implementation}
We describe the key components of the reference interpreter with the aim of illustrating the simplicity, clarity, and conciseness of the implementation. A basic knowledge of Haskell is useful but not essential to understand the content in the following sections. 

\begin{figure}[h]
\centering
\begin{tikzpicture} [align=center, arrowout]

\node(parser) at (0,0)  [box, rounded corners] {Parser};

\node(gen) at (3,0) [box, rounded corners] {Transformer};

\node(eval) at (6.5,0) [box, rounded corners] {Evaluator};

\node(apply) at (4.5,-2) [box, rounded corners] {Rule Applier};

\draw[arrowout] (-2, 0.22) -- node[above, text width=1.5cm]{\scriptsize{Graph File}} (parser.160);
\draw[arrowout] (-2, -0.22) -- node[below, text width=1.5cm]{\scriptsize{Program File}} (parser.200);

\draw [arrowout] (parser) --  (gen);
\node at (1.35,0.3) {\scriptsize{AST}};

\draw [arrowout] (gen.10) -- node[above, text width=1cm]{\scriptsize{Initial Graph}} (eval.167);
\draw [arrowout] (gen.350) --  (eval.193);
\node at (4.9,-0.5) [text width=1cm]{\scriptsize{Program}};

\draw [arrowout] (eval.325) |- (apply.350);
\node at (7.8,-1.3) {\scriptsize{Rule}};
\draw [arrowout] (eval.310) |- (apply.10);
\node at (6.4, -1.3) [text width=1cm]{\scriptsize{Graph}};
\draw[arrowout] (7,1.5) -- node[left, text width=1.5cm]{\scriptsize{Max \# Rule Apps}} (eval.45);

\draw [arrowout] (apply.90) --  (eval.225);
\node at (4.6,-1) {\scriptsize{Graphs}};

\draw[arrowout] (eval) -- node[above, text width=1cm]{\scriptsize{Output Data}} (8.5,0);

\end{tikzpicture}
\caption{Main data flow of the reference interpreter} \label{fig:architecture}
\end{figure}
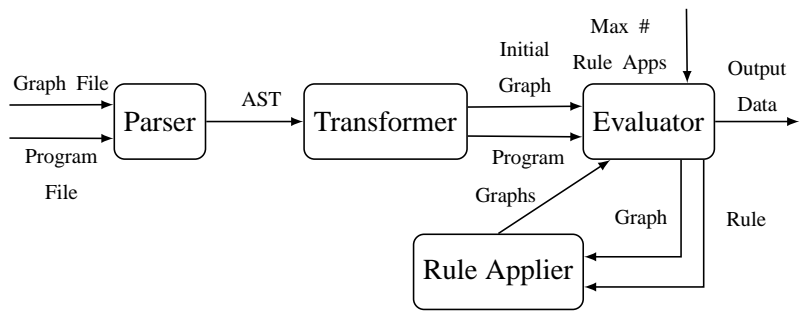

\subsection{Overview}
Figure \ref{fig:architecture} shows a data flowchart of the reference interpreter. It takes three inputs: (1) a file containing the textual representation of a GP~2 program, (2) a file containing the textual representation of a host graph, and (3) an upper limit on the number of rule applications to be made before halting program execution. It runs the program on the host graph, traversing either all nondeterministic branches of the program or a single branch, at the behest of the user. The output data is a complete description of all possible outputs. Section \ref{sec:eval} describes the output data in detail.

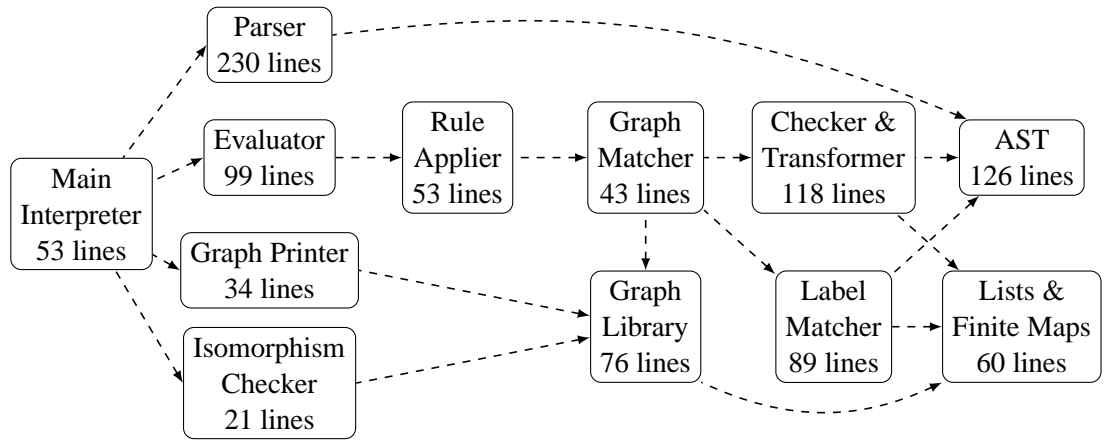
\begin{figure}
\centering
\begin{tikzpicture} [align=center]

\node(main) at (0, 0)  [box] {Main\\Interpreter\\53 lines};
\node(iso) at (2.5, -2.25) [box] {Isomorphism\\Checker\\21 lines};
\node(print) at (2.5, -0.75) [box] {Graph Printer\\34 lines};
\node(eval) at (2.5, 0.75) [box] {Evaluator\\99 lines};
\node(parser) at (2.5, 2.25) [box] {Parser\\230 lines};
\node(apply) at (5, 0.75) [box] {Rule\\Applier\\53 lines};
\node(gmatch) at (7.5, 0.75) [box] {Graph\\Matcher\\43 lines};
\node(graph) at (7.5, -1.5) [box] {Graph\\Library\\76 lines};
\node(lmatch) at (10, -1.5) [box] {Label\\Matcher\\89 lines};
\node(trans) at (10, 0.75) [box] {Checker \&\\Transformer\\118 lines};
\node(libs) at (12.5, -1.5) [box] {Lists \&\\Finite Maps\\60 lines};
\node(ast) at (12.5, 0.75) [box] {AST\\126 lines};

\draw (ast) edge[arrowin, dashed] (trans);
\draw (ast) edge[arrowin, dashed] (lmatch);
\draw (ast.145) edge[arrowin, dashed, bend right=15] (parser);
\draw (libs) edge[arrowin, dashed] (trans);
\draw (libs) edge[arrowin, dashed] (lmatch);
\draw (libs.215) edge[arrowin, dashed, bend left=25] (graph.315);
\draw (trans) edge[arrowin, dashed] (gmatch);
\draw (graph) edge[arrowin, dashed] (gmatch);
\draw (graph) edge[arrowin, dashed] (iso.0);
\draw (graph) edge[arrowin, dashed] (print.0);
\draw (lmatch) edge[arrowin, dashed] (gmatch);
\draw (gmatch.180) edge[arrowin, dashed] (apply);
\draw (apply) edge[arrowin, dashed] (eval.0);
\draw (parser.180) edge[arrowin, dashed] (main);
\draw (iso.180) edge[arrowin, dashed] (main);
\draw (eval.180) edge[arrowin, dashed] (main);
\draw (print.180) edge[arrowin, dashed] (main);

\end{tikzpicture}
\caption{Module dependencies. A module points to any modules on which it depends. Line counts exclude blank lines and comment-only lines} \label{fig:modules}
\end{figure}

The interpreter contains approximately 1,000 lines of Haskell source code. Figure \ref{fig:modules} shows the module dependency structure of the interpreter and an indication of module sizes.  

\subsection{Parser}
The parser has two components: (1) a host graph parser and (2) a program text parser. Each individual parsing function takes a string as input and attempts to match a prefix of the string to a particular syntactic unit. It uses a library of parser combinators. Their purpose is to neatly compose the parsing functions to cover standard parsing requirements such as alternation and repetition. The parsing code is very similar in appearance to GP~2's context-free grammar: each nonterminal of the grammar is represented by a Haskell function that parses the right-hand side of the grammar rule. For example:

\begin{verbatim}
gpMain :: Parser Main
gpMain = keyword "Main" |> keyword "=" |> pure Main <*> 
         commandSequence
\end{verbatim}

The operators \texttt{|> and <*>} are binary functions: \texttt{|>} ignores the output of its left parser and \texttt{<*>} sequences two parsers. Applications of \texttt{keyword} recognise and discard a string argument, and \linebreak \texttt{commandSequence} is another parsing function. \texttt{Main} is a data constructor for the main node of GP~2's abstract syntax tree.

\subsection{Checking \& Transformation}

The checking and transformation phase extracts semantic information from the AST, such as the types of variables specified in a rule schema's parameter list, and transforms both rule graphs and the host graph into the data structure defined in the graph library. The internal graph representation is a pair of maps from keys to labels for each of nodes and edges separately. Node keys are integers. Edge keys are triples: source key, target key and an integer. Node and edge labels are encoded into the node and edge data types. Operations on graphs are concisely represented using Haskell functions from the Haskell library \texttt{Data.Map} which implements maps efficiently as balanced binary trees. Node and edge enumeration functions also support the use of Haskell's strong list-processing. See Section \ref{sec:graph-match} for details.

\subsection{Label Matching}
The label matching algorithm establishes whether a label from a rule's left-hand side can be matched with a label from the host graph. It takes as input the current \textit{environment}, the set of bindings for label variables, and the two labels to be compared. 

GP~2 labels consist of a mark and a list. The marks are encoded as an abstract data type and are directly comparable. GP~2's lists are naturally encoded as Haskell lists, where each element is a GP~2 atom. Atoms occurring in the host graph are constants (integers, characters or strings), while rule atoms are either constants, variables or a concatenated string\footnote{Expressions and degree operators are forbidden in LHS labels to prevent ambiguous matching.}. If a match binds a variable, the binding must define a compatible extension of the environment.

When comparing atoms, the interesting case occurs if a list variable is encountered. GP~2 allows at most one list variable in any label expression on a left-hand side. This restriction allows binding to host-label segments of determined length, by comparing the lengths of the remainder of the rule label and the remainder of the host label. Matching fails if too few host atoms remain.

\subsection{Graph Matching}\label{sec:graph-match}

Given a rule graph $L$ and a host graph $G$, the graph matcher lazily constructs a list of \texttt{GraphMorphisms}. A \texttt{GraphMorphism} is a data structure containing an environment, a mapping between nodes in $L$ and the corresponding nodes in $G$, and a similar edge map. We use association lists to represent these small mappings, for simplicity and amenability to list-processing.  Morphisms are generated in two stages. First the candidate \texttt{NodeMorphisms} are identified, where a \texttt{NodeMorphism} is an environment and a node mapping. For each such \texttt{NodeMorphism}, the matcher searches for compatible edge mappings and environment extensions to form a set of complete \texttt{GraphMorphisms}.

\vspace{.5\baselineskip}
\noindent
\emph{Node matching.}
For each node $l_k \in L$, the matcher constructs the list of all host nodes \texttt{[$h_{k_1}, \ldots, h_{k_m}$]} that match $l_k$ with respect to label matching and rootedness\footnote{Graphs can be augmented with root nodes to reduce the search space. GP~2's semantics requires that a root node in $L$ must only match a root node in $G$ \cite{Bak-Plump12a}.} An environment is paired with each host node. The result is a list of lists \texttt{[[$h_{1_1}, \ldots, h_{1_m}$],\ldots,[$h_{n_1}, \ldots, h_{n_m}$]]} where $n$ is the number of nodes in $L$. A candidate node mapping is found by injectively selecting one item from each list. The final step is to test each candidate mapping for compatibility with respect to its environment. Haskell's list comprehensions are perfectly suited for this task: the list of lists is computed with a single nested list comprehension, while a second list comprehension is responsible for collating the valid candidate mappings. 

\vspace{.5\baselineskip}
\noindent
\emph{Edge matching.} 
For each edge in $L$, we use a candidate node morphism to determine the required source and target for a corresponding edge in the host graph. The list of candidate host edges is the list of host edges from that source to that target. Each rule edge is checked against each candidate host edge for label compatibility, supported by the environment passed from the node morphism.

\subsection{Rule Application}
Each of the \texttt{GraphMorphisms} produced by the graph matcher is checked against a \emph{dangling condition} and any rule conditions. If these checks succeed, the rule application is performed in the following steps: delete edges, delete nodes, relabel nodes, add nodes, relabel edges, add edges. For relabelling, variables take their values from a \texttt{GraphMorphism}'s environment. 

The dangling condition can be elegantly expressed as follows.
\begin{verbatim}
danglingCondition :: HostGraph -> EdgeMatches -> [NodeId] -> Bool
danglingCondition h ems delns = 
         null [e | hn <- delns, e <- incidentEdges h hn \\ rng ems]
\end{verbatim}

The second argument is an edge map, obtained from a \texttt{GraphMorphism}. The third argument is the set of nodes deleted by the rule. The function body specifies that no host edge \ttt{e} incident to any deleted node \ttt{n} may lie outside of the range of the edge map \ttt{ems}.

\subsection{The Evaluator}\label{sec:eval}
The evaluator applies a GP~2 program to a host graph, subject to an upper bound on the number of rule applications. Often the same graph can be reached through several distinct computational branches. Therefore, when program execution is complete, an isomorphism checker is used to collate the list of output graphs into its isomorphism classes. The output is as follows:

\begin{enumerate}
\item A list of unique output graphs, up to isomorphism, with a count of how many isomorphic copies of each graph were generated.
\item The number of failures. For example, a failure occurs in some contexts if none of a set of rules can be applied to a graph.
\item The number of unfinished computations. A computation is unfinished if the bound on rule applications is reached before the end of the main command sequence.
\end{enumerate}

During program execution the evaluator maintains a list of \texttt{GraphStates}, one for each nondeterministic branch of the computation so far. A \texttt{GraphState} is one of: (1) a graph with its rule application count, (2) a failure symbol with its rule application count, and (3) an unfinished symbol. Each GP~2 control construct is evaluated by a function that takes as input a single \texttt{GraphState} and some program data, returning a list of \texttt{GraphStates}. Only the application of a rule can yield a \texttt{GraphState} with a changed graph.
The rule application process is the workhorse of the interpreter, so here by way of illustration is the
top-level defining equation for the evaluation of a rule-call command:

\begin{verbatim}
evalSimpleCommand max ds (RuleCall rs) (GS g rc) = 
   if rc == max then [Unfinished]
   else case [h | r <- rs, h <- applyRule g $ ruleLookup r ds] of
        [] -> [Failure rc]
        hs -> [GS h (rc+1) | h <- hs]
\end{verbatim}

Here \texttt{max} is the rule application bound, \texttt{ds} is a list of the rule and procedure declarations in the GP~2 program, \texttt{rs} is a list of rules, and \texttt{GS g rc} is the current graph state. \texttt{GS} is the \texttt{GraphState} constructor, \texttt{g} is the working host graph, and \texttt{rc} is the number of rules that have been applied to \texttt{g}. The case-subject list comprehension can be read as, ``for all rules \texttt{r} in \texttt{rs}, apply \texttt{r} to \texttt{g} and produce the list of all output graphs \texttt{h}.'' Each individual rule application may produce multiple output graphs; the list comprehension gathers every possible output into a single lazily-computed list. If \texttt{resultGraphs} is empty, then no rule in \texttt{rs} was applicable, and the list containing the single \texttt{GraphState Failure} is returned. Otherwise, the output graphs are placed into a fresh list of \texttt{GraphStates}, each with an incremented rule-application count.

\section{Performance Evaluation}
\label{sec:performanceevaluation}

In this section we will look at how efficiently our interpreter executes the benchmark programs described in Section \ref{sec:benchmark}, and discuss the factors that affect its performance. Though not tuned for speed, the interpreter must run fast enough to allow its use as a practical tool.


\subsection{The Test Environment}

We compiled the interpreter using the Glasgow Haskell Compiler\cite{ghc} version 7.6.3 with optimisations and profiling support enabled:

\begin{verbatim}
$ ghc -O2 -prof -fprof-auto -rtsopts -o gp2 Main.hs
\end{verbatim}

All figures reported were obtained using a quad-core Intel i7 clocked at 3.4GHz, with 8GB RAM, running 64-bit Ubuntu 14.04 LTS with kernel 3.13.0. The number of processor cores should not have a significant effect on the measured performance of the single-threaded GP~2 interpreter.

We ran benchmarks using the following command

\begin{verbatim}
$ timeout --foreground 5m time \
      gp2 +RTS -p -sgc.prof -RTS $GPOPT $PROG $GRAPH 10000
\end{verbatim}

\noindent
limiting execution time to five minutes for each application of a program to a host graph. We used the sum of user and system time reported by the standard \texttt{time} utility as our measure of execution time.
The arguments to \texttt{gp2} between \texttt{+RTS} and \texttt{-RTS} tell the Haskell run-time system to save profiling information.  The \texttt{\$GPOPT} variable was either set to \texttt{--one} to put the interpreter into single-result mode (see Table \ref{table:resultsSingle}), or unset for all-result mode (see Table \ref{table:resultsAll}).
The final three mandatory arguments to the \texttt{gp2} executable specify the benchmark program, the host graph, and the maximum number of rule applications, as described in Section \ref{sec:implementation}.

\subsection{Host Graphs}
\label{subsec:hosts}

The names of host graphs used for benchmarking give an indication of their structure.

\vspace{.5\baselineskip}
\noindent
\emph{Gen $n$}. The \textit{Sierpinski} program expects a host graph containing a single node with a numeric label, which controls the number of iterations of the \texttt{expand!} command.

\vspace{.5\baselineskip}
\noindent
\emph{Linear $n$}. A chain of $n$ nodes. The first node has only a single outgoing edge. The last node has only a single incoming edge. All other nodes have exactly one incoming and one outgoing edge.

\vspace{.5\baselineskip}
\noindent
\emph{Cyclic $n$}. As Linear $n$, but with an extra edge from the last node to the first, so every node has exactly one incoming and one outgoing edge.

\vspace{.5\baselineskip}
\noindent
\emph{$x \times y$ Grid}. A rectangular lattice $x$ nodes wide by $y$ nodes tall, with $x(y-1) + y(x-1)$ edges. The \textit{shortest distances} benchmark requires all edges to have an integer ``cost'' of traversal. The \textit{grid} host graphs passed to this program have the top-left node marked grey, all edges directed either rightwards or downwards, a cost of one assigned to half of the edges, and a cost of two to the other half.

\subsection{Benchmark performance}\label{sec:benchperf}


\vspace{.5\baselineskip}
\noindent
\emph{Single-result mode.}
Table \ref{table:resultsSingle} summarises results for the reference interpreter operating in single-result mode. The \textit{Apps} column shows the number of rule applications required to reach the solution. \textit{Time} lists the sum of user and system time reported by the \texttt{time} command. The final two columns show the maximum amount of memory requested by the \texttt{gp2} executable, and the maximum memory holding live data respectively. The disparity between these two numbers, which sometimes approaches a factor of three, results from the Haskell run-time system requesting memory from the operating system in large chunks.

\begin{table}[t]
\begin{minipage}{\textwidth}
\centering

\begin{tabular}{llrrcrr}
\hline 
&  & & & & \multicolumn{2}{c}{Heap/kB}\\
Benchmark          & Host Graph & Apps & Time/s   & & Allocd & Live \\
\hline 
Acyclicity test
 &             3x3 grid &    12 &    0.02 & &  2048 &   129 \\
 &             5x5 grid &    40 &    0.03 & &  3072 &   382 \\
 &             7x7 grid &    84 &    0.17 & &  4096 &  1119 \\
 &             9x9 grid &   144 &    0.70 & &  6144 &  2100 \\
 &           cyclic 100 &     0 &    0.04 & &  3072 &   778 \\
 &           cyclic 500 &     0 &    0.46 & & 14336 &  5646 \\
 &          cyclic 1000 &     0 &    1.76 & & 25600 & 10368 \\
\hline
Shortest distances
 &             5x5 grid &    38 & $<0.01$ & &  3072 &   414 \\
 &             7x7 grid &    90 &    0.08 & &  4096 &  1177 \\
 &             9x9 grid &   175 &    0.39 & &  8192 &  3172 \\
\hline
Sierpinski
 &                gen 2 &     7 & $<0.01$ & &  2048 &   133 \\
 &                gen 3 &    17 &    0.14 & &  5120 &  1056 \\
 &                gen 4 &    45 &    6.52 & & 58368 & 18313 \\
 &                gen 5 & - & $>5m$ & & - & - \\
\hline
Transitive closure
 &            linear 05 &     6 & $<0.01$ & &  2048 &   144 \\
 &            linear 10 &    36 &    0.04 & &  2048 &   144 \\
 &            linear 20 &   171 &    1.67 & & 21504 &  7073 \\
 &            linear 30 &   406 &   14.39 & & 103424 & 33152 \\
 &            linear 40 &   741 &   66.31 & & 324608 & 103275 \\
 &            linear 50 & - & $>5m$ & & - & - \\
\hline
Vertex colouring
 &             3x3 grid &    27 &    0.02 & &  2048 &   140 \\
 &             5x5 grid &   125 &    0.03 & &  3072 &   999 \\
 &             7x7 grid &   343 &    0.17 & &  9216 &  3681 \\
 &             9x9 grid &   729 &    0.89 & & 25600 & 11438 \\
\hline

\end{tabular}

\caption[Reference interpreter benchmarks]{Reference interpreter benchmark results when generating a single output graph}

\label{table:resultsSingle}
\end{minipage}
\end{table}



\vspace{.5\baselineskip}
\noindent
\emph{All-result mode.}
Table \ref{table:resultsAll} summarises the performance of the reference interpreter running in all-result mode. This table contains three additional columns showing the total number of output graphs, the number of distinct output graphs up to isomorphism, and the number of executions that terminated in failure.
Where different solutions required differing numbers of rule applications the \textit{Apps} column now shows the range of values.

\begin{table}[t]
\begin{minipage}{\textwidth}
\centering

\begin{tabular}{llrrrrrcrr}
\hline 
&  & \multicolumn{3}{c}{Output Graphs} & & && \multicolumn{2}{c}{Heap/kB}\\
Benchmark          & Host Graph & Total & Unique   & Failed & Apps & Time/s   & & Total  & Live \\
\hline 
Acyclicity test
 &             2x2 grid &      6 &         1 &     0 &     4 & $<0.01$ & &  2048 &   134 \\
 &             3x3 grid &  19770 &         1 &     0 &    12 &   12.00 & & 10240 &  3301 \\
 &             4x4 grid & - & - & - & - & $>5m$ & & - & - \\
 &           cyclic 100 &      0 &         0 &   100 &     0 &    0.06 & &  4096 &   784 \\
 &           cyclic 500 &      0 &         0 &   500 &     0 &    0.86 & & 14336 &  5651 \\
 &          cyclic 1000 &      0 &         0 &  1000 &     0 &    3.31 & & 26624 & 11053 \\
\hline
Shortest distances
 &             2x2 grid &      6 &         1 &     0 &     4 & $<0.01$ & &  2048 &   131 \\
 &             3x3 grid &  28924 &         1 &     0 &  9-14 &   19.15 & & 167936 & 58180 \\
 &             4x4 grid & - & - & - & - & $>5m$ & & - & - \\
\hline
Sierpinski
 &                gen 2 &      6 &         1 &     0 &     7 &    0.04 & &  3072 &   242 \\
 &                gen 3 & - & - & - & - & $>5m$ & & - & - \\
\hline
Transitive closure
 &            linear 05 &    866 &         1 &     0 &     6 &    0.44 & &  6144 &  1699 \\
 &            linear 10 & - & - & - & - & $>5m$ & & - & - \\
\hline
Vertex colouring
 &             2x2 grid &    480 &         2 &     0 &   6-8 &    0.07 & &  5120 &  1598 \\
 &             3x3 grid & - & - & - & - & $>5m$ & & - & - \\
\hline

\end{tabular}

\caption[Reference interpreter benchmarks]{Reference interpreter benchmark results when generating all possible output graphs}

\label{table:resultsAll}
\end{minipage}
\end{table}

The extra costs of evaluating a program in all-result mode go beyond those of generating all possible output graphs; the interpreter must also test them for isomorphism. Unsurprisingly, execution time increases sharply with increasing size of host graph, putting many of the computations that completed in single-result mode beyond our five-minute execution-time limit.

The effect on heap usage of producing all possible results is less than one might expect for the \textit{3x3 grid} host graph in both the \textit{acyclicity test} and \textit{shortest distances} programs, given the tens of thousands of isomorphic graphs generated. We benefit from Haskell's lazy evaluation of the list of output graphs.
As there is a single isomorphism class, at most two final host graphs are needed in memory simultaneously --- though there may be many intermediate graphs awaiting further processing.

In contrast, the \textit{vertex colouring} benchmark has many distinct solutions.
As the five minute limit approached during all-results computation for the \textit{3x3 grid} host graph, \texttt{gp2} had been allocated over seven gigabytes, putting a conservative estimate of its live heap in excess of two gigabytes!

\subsection{Discussion}

In single-result mode, performance is acceptable even for some quite complex programs. However, in all-result mode, execution time and memory usage can increase very rapidly with problem size. An extreme example is the vertex-colouring program, which exhibits factorial growth in the number of possible intermediate graphs as edge-counts in initial graphs increase.

The current version of the interpreter uses a finite-map library for indexed sets of nodes and edges in graphs.
Early versions stored these sets as association lists, resulting in an interpreter which spent most of its
execution time traversing lists of nodes and edges.
The cumulative effect of several incremental improvements to our original prototype, without making it larger or
more complicated, was a large speed-up. This in turn enabled us to run larger computations,
putting greater stress on stack and heap memory.  There may yet be quite simple modifications that would reduce
memory demand --- we have made comparatively little effort in this direction.

As discussed in Section \ref{sec:graph-match} the reference interpreter matches nodes and edges in separate passes. This makes for a simple algorithm at the expense of performance. A more performance focussed implementation might use a \textit{search plan}\cite{Zundorf96, Dorr95b} in which a graph morphism is built incrementally by adding both nodes and edges to an existing partial morphism, back-tracking if no suitable candidate can be found.





\section{Related and Future Work}
\label{sec:relatedandfuture}

Early programming languages were often defined by their implementations,
perhaps in the form of a \emph{definitional interpreter}.
We now have more abstract techniques for defining operational semantics.
However, in recent years there has been a 
rehabilitation of interpreters as executable counterparts to semantic
definitions --- eg. \cite{Campbell2012}. 
Motivation varies, but here's an extract from the preface
of an influential textbook: 

\pagebreak

\begin{quote}
\textit{Our goal is to provide a deep, working understanding of the essential concepts of programming languages. \ldots
Most of these essentials relate to the semantics, or meaning, of program elements. Such meanings reflect how program elements are interpreted as the program executes. \ldots
The most interesting question about a program is, \textnormal{``What does it do?''} The study of interpreters tells us this. Interpreters are critical because they reveal nuances of meaning, and are the direct path to more efficient compilation and to other kinds of program analyses.} \cite{Friedmanetal2008}
\end{quote}
In several respects, our motivation is similar.
We adopt the slogan: \emph{Semantics first!}.
But then, following the semantic definition, we write a reference interpreter in order to
promote a ``\textit{deep, working understanding}'' of the GP~2 design,
and to find ``\textit{path(s) to more efficient compilation \ldots and program analysis}''.

Languages based on graph-transformation rules include
\textsc{Progres} \cite{Schuerr-Winter-Zuendorf99a},
\textsc{Agg} \cite{Ermel-Rudolf-Taentzer99a,Runge-Ermel-Taentzer11a},
\textsc{Gamma} \cite{Fradet-LeMetayer98a},
\textsc{Groove} \cite{Ghamarian-deMol-Rensink-Zambon-Zimakova12a},
\textsc{GrGen.Net} \cite{Jakumeit-Buchwald-Kroll10a} and
\textsc{Porgy} \cite{Fernandez-Kirchner-Mackie-Pinaud14a}.
To our knowledge, none of these languages has a published implementation in the same spirit as our reference interpreter. For example, \textsc{Groove} and \textsc{GrGen.Net} are two of the most widely used systems. The Java source code for the \textsc{Groove} implementation, including a graphical development suite, extends to around 150,000 lines. \textsc{GrGen.Net} is implemented in a combination of Java and C\#: a Java front-end is used to generate C\# code and .NET assemblies from a textual specification of a \textsc{GrGen} program; the run-time system and other components are written in C\#. In all there are around 68,000 lines of Java source for the front-end, and around 93,000 lines of C\# for the run-time system, API support and an interactive shell.
We recognise that both \textsc{Groove} and \textsc{GrGen.Net} are mature and fully-featured systems, and \textsc{GrGen.Net} in particular is highly optimising. Even so, the contrast with the 1,000-line Haskell sources for our GP 2 reference interpreter is striking.

We have begun work on two compiled implementations of GP~2. One generates code for an abstract machine; the other translates GP~2 programs to C. They also differ in the way a low-level graph data structure is defined and accessed, and the strategies employed to match left-hand sides of rules. The reference interpreter is supporting these ongoing developments. For example, some front-end components are re-used, and we check output graphs against isomorphism classes
computed by the interpreter.

\section{Conclusions}
\label{sec:conclusions}


Our original goals for our reference interpreter have largely been realised. We have a concise implementation of GP~2, expressed in around 1,000 lines using the lazy functional language Haskell. We have taken every opportunity to use a Haskell strength --- lazy list-processing, and in particular list comprehensions for generate-and-test style definitions --- to achieve this conciseness. However, despite our observations in Section~\ref{sec:usesrequirements} about error reports and traces, we concede that our current interpreter provides only a bare minimum in this respect.

As stated in the Introduction, our motivation for producing a simple interpreter was to achieve clarity and correctness. This raises the question of whether the reference interpreter could be formally verified against the operational semantics of GP~2. While this is a desirable goal for future work, existing verification projects for subsets of C \cite{Leroy09a} and ML \cite{Kumar-Myreen-Norrish-Owens14a} indicate that such a project would be a major endeavour despite the modest size of the GP~2 language.

When working with the interpreter, we have had some unexpected results. Occasionally, the practical consequences of a crisp semantic definition may be surprising to programmers, or it may pose challenges for an efficient implementation. We have found that our reference interpreter can shed helpful light in such instances.

As we have shown in Section~\ref{sec:performanceevaluation}, the interpreter is efficient enough for practical use in testing,
both by GP~2 programmers and by the developers of other GP~2 implementations.
Our main reservation here concerns all-results mode.
Used in this mode, the interpreter can require very long execution times and all the memory our machines have available.
One remedy might be to check for isomorphism or other equivalences
between intermediate graphs, compacting the state-space.
However, the extra machinery would complicate the interpreter, and it
could demand even more space in some cases.
Instead, our likely solution will be to build up a standard set of test programs.
We can first run each test (for several days, if necessary)
on a powerful machine to produce the set of all possible output graphs up to isomorphism.
Our isomorphism checker, though simple, is efficient enough for rapid
subsequent checking of single results produced by another implementation.

\vspace{\baselineskip}
\noindent\textbf{Acknowledgements.} We are grateful to Berthold Hoffmann and the anonymous referees for their comments which helped to improve the presentation of this paper.

\bibliography{abbr,bibliography}{}
\bibliographystyle{eptcs}
\end{document}